\documentclass[preprint,aps,tightenlines,showpacs,superscriptaddress]{revtex4}
\usepackage[dvips]{graphicx}
\usepackage{dcolumn}
\usepackage{bm}
\usepackage{epsfig}

\newcommand{\beq}{\begin{equation}}
\newcommand{\eeq}{\end{equation}}

\begin{document}
\title{On the Methods for Constructing Meson-Baryon Reaction Models within
Relativistic Quantum Field Theory\footnote{Notice: Authored 
by Jefferson Science Associates, LLC under U.S. DOE Contract No. DE-AC05-06OR23177. 
The U.S. Government retains a non-exclusive, paid-up, irrevocable, world-wide 
license to publish or reproduce this manuscript for U.S. Government purposes. 
}} 

\vspace{0.5cm}
\author{B. Juli\'a-D\'{\i}az}
\affiliation{ Excited Baryon Analysis Center (EBAC), Thomas Jefferson National
Accelerator Facility, Newport News, VA 23606, USA}
\affiliation{Department d'Estructura i Constituents de la Mat\`{e}ria
and Institut de Ci\`{e}ncies del Cosmos,
Universitat de Barcelona, E--08028 Barcelona, Spain}
\author{H. Kamano}
\affiliation{ Excited Baryon Analysis Center (EBAC), Thomas Jefferson National
Accelerator Facility, Newport News, VA 23606, USA}
\author{T.-S. H. Lee}
\affiliation{ Excited Baryon Analysis Center (EBAC), Thomas Jefferson National
Accelerator Facility, Newport News, VA 23606, USA}
\affiliation{Physics Division, Argonne National Laboratory, 
Argonne, IL 60439, USA}
\author{A. Matsuyama}
\affiliation{ Excited Baryon Analysis Center (EBAC), Thomas Jefferson National
Accelerator Facility, Newport News, VA 23606, USA}
\affiliation{Department of Physics, Shizuoka University, Shizuoka 422-8529, Japan}
\author{T. Sato}
\affiliation{ Excited Baryon Analysis Center (EBAC), Thomas Jefferson National
Accelerator Facility, Newport News, VA 23606, USA}
\affiliation{Department of Physics, Osaka University, Toyonaka, 
Osaka 560-0043, Japan}
\author{N. Suzuki}
\affiliation{ Excited Baryon Analysis Center (EBAC), Thomas Jefferson National
Accelerator Facility, Newport News, VA 23606, USA}
\affiliation{Department of Physics, Osaka University, Toyonaka,
Osaka 560-0043, Japan}

\begin{abstract}
Within the relativistic quantum field theory,
we analyze the differences between the $\pi N$ reaction models
constructed from using (1) 
 three-dimensional reductions of Bethe-Salpeter Equation, (2) 
method of unitary
transformation, and (3)  time-ordered perturbation theory.
Their relations with the approach based on the dispersion relations of S-matrix
theory are dicusssed. 
\end{abstract}
\maketitle

\section{Introduction}

Because of the non-perturbative nature of the problem and the complexities
of the reaction mechanisms, the calculations of meson-baryon reactions
within the framework of relativistic quantum field theory can not be
done exactly. Thus the progress we can make now and in the forseeable
future is to construct manageable reaction models for analyzing the data.
Each model involves some approximations and assumptions.
For understanding the information extracted from the data,
such as the electromagnetic form factors of the $N$-$N^*$ transitions, it is
necessary to examine this phenomenological aspect of the employed models.
This is the objective of this paper concerning the approaches based on
(1) three-dimensional reductions~\cite{klein}, 
(2) method of unitary transformation ~\cite{sl96,sko,msl07}, and
(3) time-ordered  perturbation theory~\cite{julich90,julich00}.
For simplicity, we will only consider the case of
single-channel $\pi N$ scattering. This is sufficient to reveal the
differences between these approaches. We will also examine the relations 
between these approaches and the
approach based on the dispersion relations of the S-matrix theory.
As is well documented~\cite{gw}, these two theoretical frameworks are based on
very different theoretical considerations and hence there is no compelling
theoretical reasons to favor one of them in developing phenomenological
models to analyze the data.

The considered three methods for constructing $\pi N$ models are explained
and analyzed in sections II-IV. In each section we will also
address their relations with the approach based on dispersion relations.
In section 5, we give a summary.

\section{Three-dimensional reductions}

To illustrate the derivations of three-dimensional equations
for $\pi N$ scattering from relativistic quantum field theory, it
is sufficient to consider a simple $\pi NN$ interaction
Lagrangian density
\begin{eqnarray}
 L_{int}(x) = \bar{\psi}(x) \Gamma_0 \psi(x)\phi(x)\, ,
\label{eq:barel}
\end{eqnarray}
where $\psi(x)$ and $\phi(x)$ denote respectively the nucleon and pion fields and
$\Gamma_0$ is a bare $\pi NN$ vertex, such as $\Gamma_0 = i g \gamma_5$ in the familiar
pseudo-scalar coupling. By using the standard method~\cite{zuber}, it is straightforward
to derive from Eq.~(\ref{eq:barel}) the Bethe-Salpeter equation for $\pi N$ 
scattering and the one-nucleon propagator. In momentum space, the resulting 
Bethe-Salpeter equation can be written as
\begin{equation}
T(k^\prime,k;P) = B(k^\prime ,k;P) + \int d^4k^{\prime\prime}
B(k^\prime , k^{\prime\prime};P)G(k^{\prime\prime};P)T(k^{\prime\prime},k;P)\;,
\label{eq:BSeq}
\end{equation}
where $k$ and $P$ are respectively the relative and total momenta
defined by the nucleon momentum $p$ and pion momentum $q$
\begin{eqnarray}
P&=&p+q \, , \nonumber \\
k&=&\eta_\pi(y)p-\eta_N(y)q \, .\nonumber
\end{eqnarray}
Here $\eta_N(y)$ and $\eta_\pi(y)$ can be any function of a chosen parameter $y$ with
the condition
\begin{eqnarray}
\eta_\pi(y) + \eta_N(y) = 1 \, .
\label{eq:choice}
\end{eqnarray}
Obviously we have from  the above definitions that
\begin{eqnarray}
p&=&\eta_N(y)P+k \, , \nonumber \\
q&=&\eta_\pi (y)P-k \, .
\end{eqnarray}
In analogy to the nonrelativistic form, they are often chosen as: 
$\eta_N = m_N/(m_\pi+m_N)$ and $\eta_\pi=m_\pi/(m_\pi+m_N)$. The choice 
of the $\eta'$s is irrelevant to the derivation presented below in 
this section provided that Eq.~(\ref{eq:choice}) is satisfied.

Note that $T$ in Eq.~(\ref{eq:BSeq}) 
is the "amputated" invariant amplitude and is
related to the $\pi N$ S-matrix by
$S \propto \bar{u}T u$ with $u$ denoting the nucleon spinor.
The driving term $B$ in Eq.~(\ref{eq:BSeq}) is the sum of all 
two-particle irreducible amplitudes, and $G$ is the product of 
the pion propagator $D_{\pi}(q)$ and the nucleon propagator $S_N(p)$.
In the low energy region, we neglect the dressing of pion propagator
and simply set
\begin{eqnarray}
D_\pi(q) =\frac{1}{q^2-m_\pi^2+i\epsilon} \, ,
\end{eqnarray}
where $m_\pi$ is the physical pion mass.
The nucleon propagator
can be written as
\begin{eqnarray}
S_N(p) = \frac{1}
{i p \hspace{-0.08in} / - m^0_N -\tilde\Sigma_N(p^2)+i\epsilon} \, ,
\label{eq:n-prop}
\end{eqnarray}
where $m^0_N$ is the bare nucleon mass and the nucleon self energy operator
$\tilde\Sigma_N$ is defined by
\begin{eqnarray}
\tilde\Sigma_N(p^2) =  \int d^4k \, \Gamma_0 \,G(k;p) \,\tilde\Gamma(k;p) \, . \label{nucl1}
\label{eq:sigmatilde}
\end{eqnarray}
The dressed vertex function $\tilde\Gamma$ on the right hand side of Eq.
(\ref{nucl1}) depends on the $\pi N$ Bethe-Salpeter amplitude
\begin{eqnarray}
\tilde\Gamma(k;P) = \Gamma_0 
+ \int d^4k^\prime \, \Gamma_0 \, G(k^\prime;P) \, T(k^\prime,k;P) \,. 
\label{nucl2}
\end{eqnarray}

It is only possible in practice
to consider the leading term of
 $B$ of Eq.~(\ref{eq:BSeq}). For the  simple
Lagrangian Eq.~(\ref{eq:barel}) the
 leading term  consists of
the direct and crossed $N$ diagrams
\begin{eqnarray}
B (k,k';P) = B^{(a)}(k,k';P) + B^{(b)}(k,k';P) \, ,
\label{eq:vterm}
\end{eqnarray}
where
\begin{eqnarray}
B^{(a)}(k,k';P) &=& \Gamma_0\frac{1}{i\gamma\cdot P - m^0_N +i\epsilon} \Gamma_0,
\label{eq:vterm-s} \\
 B^{(b)}(k,k';P) &=& \Gamma_0\frac{1}{i \gamma \cdot \bar{P}  - m^0_N +i\epsilon}\Gamma_0,
\label{eq:vterm-u}
\end{eqnarray}
with {\it $\bar P =[ {\eta_N(y)-\eta_\pi(y)}]P+k+k'$.}

Equations~(\ref{eq:BSeq})-(\ref{eq:vterm-u}) 
form a closed set of coupled equations
for determining
the dressed nucleon propagator of Eq.~(\ref{eq:n-prop}) and the
$\pi N$ Bethe-Salpeter amplitude of Eq.~(\ref{eq:BSeq}). It is important to
note here that this is a drastic simplification of the original
field theoretical problem defined by the Lagrangian~(\ref{eq:barel}).
However, it is still very difficult to
solve this highly nonlinear problem exactly.
For practical applications further approximations 
are usually introduced.

The first step is to define the physical
nucleon mass by imposing the condition that
the dressed nucleon propagator
should have the limit
\begin{eqnarray}
S_N(p)|_{p^2\rightarrow m^2_N}
 \rightarrow \frac{1}{i p \hspace{-0.08in} / - m_N+i\epsilon} \, ,
\end{eqnarray}
where  $m_N$ is the physical nucleon mass.
This means that the self-energy in the nucleon propagator Eq.~(\ref{eq:n-prop})
is constrained by the condition
\begin{eqnarray}
m^0_N + \tilde\Sigma_N(m_N^2) = m_N.
\label{eq:n-mass}
\end{eqnarray}
The next step is to assume that the $p$-dependence of the nucleon self-energy is weak
and we can use the condition Eq.~(\ref{eq:n-mass}) 
to set $m^0_N + \tilde\Sigma(p^2)\sim m^0_N
+\tilde\Sigma(m_N^2) = m_N$. This approximation greatly simplifies the nonlinearity of
the problem, since the full $\pi N$ propagator $G$ in Eqs. (2), (7) and (8)
then takes
the following simple form
\begin{eqnarray}
G(k;P)=\frac{1}{i p \hspace{-0.08in} / - m_N+i\epsilon}
\,\frac{1}{q^2-m^2_\pi+i\epsilon}.
\end{eqnarray}

The next commonly used approximation is to reduce the
dimensionality of the above integral equations from four to three.
The resulting models will be called 3dBS models in the following discussions.
There exists extensive literature on this subject, as reviewed in
Ref.~\cite{klein}. A rather complete numerical study of several of 
these 3dBS $\pi N$ models was given in Ref.~\cite{hung}. We therefore will 
not get into these subjects. Instead we will focus on how these models are 
interpreted in the actual analysis of $\pi N$ data.

Let us specifically consider a 3dBS model derived from using the
three-dimensional reduction method of Kadyshevsky~\cite{kady}. In 
the c.m. system, $P=(E,\vec{0})$, we first define
\begin{eqnarray}
t(\vec{k}^{\,\prime},\vec{k},E)&=&
\frac{1}{(2\pi)^3}\sqrt{\frac{m_N}{E_N(k')}}\frac{1}{\sqrt{2E_\pi(k')}}
\,
\bar{u}_{\vec{k'}}\,
T(\vec{k'},\vec{k},E)\,
u_{\vec{k}} \,
\sqrt{\frac{m_N}{E_N(k)}}\frac{1}{\sqrt{2E_\pi(k)}} \\
v(\vec{k}^{\,\prime},\vec{k},E)&=&
\frac{1}{(2\pi)^3}\sqrt{\frac{m_N}{E_N(k')}}\frac{1}{\sqrt{2E_\pi(k')}}
\,\bar{u}_{\vec{k}}
\,B(\vec{k'},\vec{k},E)
\,u_{\vec{k}}
\, 
\sqrt{\frac{m_N}{E_N(k)}}\frac{1}{\sqrt{2E_\pi(k)}}
\end{eqnarray}
where the time components of the momenta of the BS amplitude $T$ and
interaction $B$ in Eq.~(\ref{eq:BSeq}) 
have been fixed by the chosen three-dimensional reduction,
as explicitly given in Ref.~\cite{hung}.
The resulting three-dimensional scattering equation can then be cast into
the following form 
\begin{eqnarray}
t(\vec{k}^{\,\prime},\vec{k},E) = v(\vec{k}^{\,\prime},\vec{k},E)
+\int d\vec{k}^{\,\prime\prime} 
v(\vec{k}^{\,\prime}, \vec{k}^{\,\,\prime\prime},E) 
g({k}^{\,\prime\prime},E)
t(\vec{k}^{\,\prime\prime},\vec{k},E) 
\label{eq:3dBS-eq}
\end{eqnarray}
where the propagator is 
\begin{eqnarray}
g(k,E)=\frac{1}{E-E_N(k)-E_\pi(k)+i\epsilon}\,.
\label{eq:3dBS-eq-p}
\end{eqnarray}
For simplicity, we only consider the case that the
interaction $v$ is derived from
$s$- and $u$-channel mechanisms defined by Eqs.(\ref{eq:vterm})-(\ref{eq:vterm-u}).
The difference between  3dBS models whose  scattering
equation can be cast into the form of Eq.~(\ref{eq:3dBS-eq}) is  
in the expression of the driving term $v$ in 
Eq.~(\ref{eq:3dBS-eq}). It is important to note that the only common condition
these 3dBS models satisfy exactly is the unitarity condition, as explained in
Ref.~\cite{klein}. 
Thus it is not surprising
that their amplitudes have different analytic structure in complex-E plane.

To simplify the presentation, we use the following operator form for Eq.
(\ref{eq:3dBS-eq})
\begin{eqnarray}
t(E) = v(E) + v(E) g(E) t(E)
\label{eq:3dbs-eqo}
\end{eqnarray}
For most of the 3dBS models, one gets the following form of $v(E)$ 
in the $P_{11}$ partial wave
\begin{eqnarray}
v(E) = v^{pole}(E) + v^{bg}(E)
\label{eq:3dbs-v}
\end{eqnarray}
with 
\begin{eqnarray}
v^{pole}(E)=\Gamma_0^\dagger\frac{1}{E-m_N^0} \Gamma_0
\label{eq:v-pole}
\end{eqnarray}
where $m^0_N$ is the bare mass of the starting Lagrangian. The form
of the second term $v^{bg}(E)$ of Eq.(\ref{eq:3dbs-v}) is 
 not important in the  the following
discussions.

Substituting Eqs.~(\ref{eq:3dbs-v})-(\ref{eq:v-pole}) 
into Eq.~(\ref{eq:3dbs-eqo}), we then get the following well known form
\begin{eqnarray}
t(E)= t^{bg}(E) +
\frac{\Gamma^\dagger(E) \Gamma(E)}{E
-m^0_N - \Sigma_N(E)},
\label{eq:p11-t}
\end{eqnarray}
where 
\begin{eqnarray}
t^{bg}(E)&=& v^{bg}(E) + v^{bg}(E) g(E)t^{bg}(E), \nonumber \\
& & \nonumber \\
 \Gamma(E)&=& \Gamma_0[ 1 + g(E)
t^{bg}(E)], \label{Gamma} \nonumber \\
& & \nonumber \\
 \Sigma_N(E) &=& \langle N_0|\Gamma_0 g(E)\Gamma^\dagger(E)|N_0\rangle .
\end{eqnarray}
In the above equations $|N_0\rangle $ is the bare one-nucleon state,
$\Sigma_N(E)$ and $\Gamma$ are the three-dimensional forms of 
Eqs~(\ref{eq:sigmatilde}),~(\ref{nucl2}), respectively. The condition 
Eq.~(\ref{eq:n-mass}) then becomes
\begin{eqnarray}
m^0_N + \Sigma_N(m_N) = m_N
\end{eqnarray}
which defines the bare mass $m^0_N$ from the physical mass $m_N$.
It is common~\cite{afnan} to use Eq.~(\ref{eq:p11-t})
at the $E\rightarrow m_N$ position to
define the dressed vertex $\Gamma (E)$ in terms of physical coupling constant.
To get this relation,
we first note that
at $E \rightarrow m_N$, the self-energy $\Sigma_N(E)$
 can be expanded as
\begin{eqnarray}
\Sigma_N(E)=\Sigma_N(m_N)+(E-m_N)\Sigma_1(m_N)
+ \cdot\cdot\cdot
\label{eq:sigm}
\end{eqnarray}
where
\begin{equation}
\Sigma_1(m_N)= \left. \frac{\partial \Sigma_N(E)}{\partial E}  \right
|_{E=m_N}\,.
\label{eq:sigm1}
\end{equation}
By using the above relations,  
Eq.~(\ref{eq:p11-t}) becomes
\begin{eqnarray}
t(E)|_{E\rightarrow m_N} = t^{bg}(m_N) + [Z^{1/2}_2\Gamma^\dagger(m_N)]
\frac{1}{E-m_N}[Z^{1/2}_2\Gamma(m_N)]
\end{eqnarray}
with 
\begin{eqnarray}
Z_2^{-1}=1-\Sigma_1(m_N).
\end{eqnarray}
The renormalized
vertex $[Z^{1/2}_2\Gamma(m_N)]$ is then used to define  the physical
coupling.
If the bare vertex $\Gamma_0$ is chosen to be
the usual pseudo-vector coupling $L_I
=\frac{f^{(0)}_{\pi NN}}{m_\pi} \bar{\psi} \gamma_5\gamma_\mu\psi \partial^\mu\phi$, the above
procedures relate the
renormalized coupling constant $f_{\pi NN}$ to the bare
coupling constant $f_{\pi NN}^{(0)}$, 
\begin{eqnarray}
f_{\pi NN} = f_{\pi NN}^{(0)} [ 1 + g(m_N) t^{bg}(m_N)]Z_2^{1/2}.
\end{eqnarray}
The renormalized coupling constant is identified with the empirical value
 $g^2_{\pi NN}/4\pi=(2m_N/m_\pi)^2(f_{\pi NN}^2/4\pi)=
14.3$. 

Here we point out that the above procedure implies an interpretation 
where the physical nucleon is made of a bare core $N_0$ and pion cloud. 
To illustrate this, it is sufficient to consider the case when $t^{(bg)}=0$. 
If we set $\langle N_0|\Gamma_0|k\rangle =\Gamma_0(k)$, Eq.~(\ref{eq:p11-t}) 
has the following analytic form, 
\begin{eqnarray}
t(k',k,E)=\Gamma^*_0(k)\frac{1}{E-m^0_N-\Sigma^0_N(E)}\Gamma_0(k)
\label{eq:t0}
\end{eqnarray}
where
\begin{eqnarray}
\Sigma^0_N(E)&=&\int k^2dk \,\frac{|\Gamma_0(k)|^2}{E- E_\pi(k)-E_N(k)+i\epsilon}
\label{eq:sigma-0} \\
\Sigma^0_1(E)&=&\frac{\partial \Sigma^0_N(E)}{\partial E} \nonumber \\
&= &- \int k^2dk\,\frac{|\Gamma_0(k)|^2}{(E- E_\pi(k)-E_N(k))^2}\,.
\end{eqnarray}
The nucleon pole condition Eq.(24) then becomes
\begin{eqnarray}
& &m_N=m^0_N+\Sigma^0_N(m_N) \label{eq:pole}\\
& & \nonumber \\
& &\langle k'|t(E)|k\rangle |_{E\rightarrow m_N} = 
\bar{\Gamma}_0^*(k') \; \frac{1}{E-m_N}\;\bar{\Gamma}_0(k)
\end{eqnarray}
where  the renormalized vertex function is
\begin{eqnarray}
\bar{\Gamma}_0(k)=Z_0^{1/2}\Gamma_0(k)
\label{eq:renf}
\end{eqnarray}
with
\begin{eqnarray}
Z_0^{-1}&=&1-\Sigma^0_1(m_N)\\
&=&1+\int dk k^2\,\frac{|\Gamma_0(k)|^2}{(m_N- E_\pi(k)-E_N(k))^2}\,.
\label{eq:z0}
\end{eqnarray} 
It is interesting to note that the above nucleon pole conditions can be
reproduced by assuming that the structure of the nucleon can be described by
the following mass operator $h$ in a subspace spanned by the state $|N_0\rangle $
and $|k\rangle $ for the $ \pi N$ state
\begin{eqnarray}
& &h=h_0 + \Gamma_0
\label{eq:nbs-h}
\end{eqnarray}
with
\begin{eqnarray}
& & h_0|N_0\rangle  = m^0_N |N_0\rangle  
\label{eq:nbs-n0} \\
& &h_0|k\rangle  = (E_N(k)+E_\pi(k)) |k\rangle  \,.
\label{eq:nbs-pin}
\end{eqnarray}
We assume that the physical nucleon state is defined by
\begin{eqnarray}
 h|N\rangle  & =& m_N|N\rangle  \label{eq:nbs-eq}, \\
& & \nonumber \\
|N\rangle  &=& z_0^{1/2}\left[|N_0\rangle  + \int k^2 dk f(k) | k\rangle \right].
\label{eq:nbd-wf}
\end{eqnarray}
The basis states are  normalized as $\langle N_0|N_0\rangle =1$, 
$\langle k|k'\rangle = k^2 \delta(k-k')$. Projecting Eq.~(\ref{eq:nbs-eq}) from the left onto
$|N_0\rangle $ and $|k\rangle $, we then get
\begin{eqnarray}
& &m_0 +\int k^2 dk f(k)\Gamma_0(k) = m_N \label{eq:nm-1}\\
& & \nonumber \\
& &[E_N(k)+E_\pi(k)] f(k) +\Gamma^*_0(k) = m_Nf(k) \label{eq:nm-2} 
\end{eqnarray}
The normalization condition $\langle N|N\rangle =1$ gives
\begin{eqnarray}
z_0=1+\int k^2dk |f(k)|^2 \label{eq:nm-3}
\end{eqnarray}
From Eq.~(\ref{eq:nm-2}) we have the solution
\begin{eqnarray}
f(k)=\frac{\Gamma^*_0(k)}{m_N -E_N(k)-E_\pi(k)}\,.
\label{eq:fk0}
\end{eqnarray}
Substituting Eq.~(\ref{eq:fk0}) into Eq.~(\ref{eq:nm-1}), we then get exactly 
the nucleon pole condition Eq.~(\ref{eq:pole}). 
Within the model defined by the mass operator Eq.~(\ref{eq:nbs-eq}),
the physical $\pi NN$ vertex can be calculated from using
Eq.~(\ref{eq:nbd-wf}) 
\begin{eqnarray}
\Gamma_N(k) =\langle N|\Gamma_0|k\rangle  = z_0^{1/2} \Gamma_0(k)
\end{eqnarray}
Substituting Eq.~(\ref{eq:fk0}) into Eq.~(\ref{eq:nm-3}), we find that
 $z_0$ is exactly
equal to $Z_0$ of Eq.~(\ref{eq:z0}) and hence $\Gamma_N(k)$ is exactly 
the renormalized vertex $\bar{\Gamma}_0$ of Eq.~(\ref{eq:renf}). 

The above simple model illustrated that  the nucleon pole 
conditions Eqs~(33)-(37) can be related to the substructure of
the nucleon.
If we write  Eq.~(\ref{eq:nbd-wf}) as an operator form
$|N\rangle =|N_0\rangle  + f|\pi N\rangle $ and iterate it, we then get
\begin{eqnarray}
|N\rangle =z^{1/2}_0[|N_0\rangle  +  f|\pi  N_0\rangle  +
f f|\pi\pi N_0\rangle  + f f f|\pi\pi\pi N_0\rangle  +\cdots \,.
\end{eqnarray}
This illustrates that
the usual procedure of requiring the $\pi N$ amplitudes to have a nucleon pole
implies that the physical nucleon is made of a $N_0$ core and meson cloud.

To be consistent, one in principle should also replace $E_N(k)$ in the
propagator Eq.~(\ref{eq:3dBS-eq-p}) by an expression
 which is related to bare mass $m^0_N$ and 
 the matrix element
of the self energy $\Sigma^0_N$, defined in Eq.~(\ref{eq:sigma-0}).
But this complicates the unitarity condition for the scattering
amplitude $t$ defined by Eq.~(\ref{eq:3dBS-eq}). To do it properly, 
one needs to also consider the $\pi\pi N$ unitarity condition since 
the self energy $\Sigma^0_N$ contains $\pi N$ intermediate state.

Let us stress that the above observation has a connection with
 the $\pi N$ scattering equation derived by 
 Aaron and Amado and Young~\cite{aay} (AAY) using the three-dimensional
reduction of Blankenbecler and Sugar~\cite{bs}. The essential assumption
of their derivation is that the $\pi N$ scattering can be described from
the isobar model where the pion is scattered from  an isobar   
system which can decay into $\pi N$. If we identify their isobar as $N_0$
of the simple mass operator defined by Eq.~(\ref{eq:nbs-h}), their equation 
can be schematically cast into the following from
\begin{eqnarray}
t(\vec{k}^{\,\prime},\vec{k},E) = v^{(opex)}(\vec{k}^{\,\prime},\vec{k},E)
+\int d\vec{k}^{\,\prime\prime}
v^{(opex)}(\vec{k}^{\,\prime}, \vec{k}^{\,\,\prime\prime},E)
g_{AAY}({k}^{\,\prime\prime},E)
t(\vec{k}^{\,\prime\prime},\vec{k},E)
\label{eq:aay-eq}
\end{eqnarray}
where the propagator is
\begin{eqnarray}
g_{AAY}(k,E)=\frac{1}{E-E_{N_0}(k)-E_\pi(k) - \Sigma^0_N(k,E) +i\epsilon}\,.
\label{eq:aay-p}
\end{eqnarray}
Here $E_{N_0}(k) =[(m^0_N)^2+k^2]^{1/2}$ and
$\Sigma^0_N(k,E)$ is determined by properly boosting the expression
Eq.~(\ref{eq:sigma-0}). 
The driving term of Eq.~(\ref{eq:aay-eq}) is determined by the
one-nucleon-exchange mechanism
\begin{eqnarray}
v^{(opex)}(\vec{k}',\vec{k},E) =[\Gamma^*_0(k)Z_0^{1/2}]\frac{1}
{E-E_N(\vec{k}+\vec{k}')-E_\pi(k)-E_\pi(k')+i\epsilon}
[\Gamma_0(k')Z_0^{1/2}]\,.
\end{eqnarray}
Note that $\Sigma^0_N(k,E)$ and $v^{(opex)}$ in the above equation are defined
by the same vertex function $\Gamma_0(k)$.
In the AAY approach, this consistent treatment of
the propagator $g_{AAY}(k,E)$ and the interaction
$v^{(opex)}$ is the consequence of requiring that the scattering
equation satisfies the $\pi\pi N$ unitarity condition.
In other words, if one replaces the propagator $g_{AAY}(k,E)$ by 
$g(k,E)$ of Eq.~(\ref{eq:3dBS-eq-p}), the resulting amplitude 
from solving Eq.~(\ref{eq:aay-eq}) will not satisfy the 
unitarity condition. With some derivations, one can also see 
that the solution of Eq.~(\ref{eq:aay-eq}) will not have a 
pole at $E=m_N$. The AYY approach simply was not developed to 
reproduce the same analytic structure of the dispersion 
relations in the unphysical region $E \leq m_\pi+m_N$.

To further explore the differences with the approaches based on 
 dispersion relations, we note that the models considered in this
section as well as in the next one solve integral equations, 
such as Eq.~(\ref{eq:3dBS-eq}), and require form factors to 
regularize the matrix elements of the potential $v$ and the 
vertex interaction $\Gamma_0$. These form factors can give 
poles to the on-shell scattering amplitudes in the unphysical 
region of $E\leq m_N+m_\pi$ where the nucleon pole is identified. 
More explicitly, if a dipole form is used to parameterize $\Gamma_0$,
the on-shell matrix element of Eq.~(\ref{eq:t0}) becomes
\begin{eqnarray}
t(k_0,k_0,E) \sim 
\left[ \frac{\Lambda^2}{k^{ 2}_0+\Lambda^2} \right]^2
\frac{1}{E-m_0 - \Sigma^0_N(E)} 
\left[\frac{\Lambda^2}{k_0^2+\Lambda^2}\right]^2
\label{eq:p11-t-a}
\end{eqnarray}
where $k_0$ is defined by $E=E_N(k_0) + E_\pi(k_0)$. Thus this amplitude
can have pole in the region where we define the nucleon pole if 
$\Lambda \leq m_N+m_\pi$. This illustrates that the analytic structure of 
the dynamical models deduced from relativistic quantum field theory can 
not be completely consistent with that defined by the dispersion relations
of the S-matrix theory. In fact, there is no compelling reason to require
that they have the same analytic structure.  In the very extensive literature, 
as thoroughly reviewed in the textbook of Goldberger and Watson~\cite{gw}, 
the widely used fixed-$t$ dispersion relations in analyzing $\pi N$ 
scattering can not be derived from relativistic quantum field theory 
$exactly$. Historically, the S-matrix theory is considered as an 
alternative to relativistic quantum field theory to study strong
interactions. There is no rigorous theoretical argument to favor one of them 
in developing phenomenological models to analyze the data.

\section{Method of Unitary Transformation}

The method of unitary transformation was essentially based on the same idea
of the Foldy-Wouthuysen transformation developed in the 
study of electromagnetic
interactions. Instead of considering the original Lagrangians with bare masses
and bare vertex interactions, we simply ask how the strong interactions
can be described with a phenomenological 
Lagrangian defined by the physical masses and physical coupling constants.
It is understood that the application of such a phenomenological Lagrangian
to calculate any amplitude should drop loops 
associated with one-particle
states and vertices which are already absorbed in the
definitions of physical masses and coupling constants.
It is an non-trivial problem to justify these rules within the exact theory.
But such rules are valid in practice since we will only consider
leading order terms of a perturbative expansion which will be specified later.
This means that we assume that we have already solved the one-particle problem
within a model, such as that defined by 
Eqs.~(\ref{eq:nbs-h})-(\ref{eq:nbd-wf}),
and this problem will not be dealt with in developing reaction models.
This is the main difference between the models based on three-dimensional
reductions described in section II and the model based on the
unitary transformation. The advantage of the latter is that the unitarity condition can be
satisfied trivially; in particular in handling the multi-channel multi-resonance
reactions. Of course, the price we pay is that the connection to
the theory of nucleon structure is perhaps more remote than the approaches 
based on three dimensional reductions.

To illustrate the method of unitary transformation, 
we again consider the simplest phenomenological
Lagrangian density
\begin{eqnarray}
{\it L}(x)={\it L}_0(x) + {\it L}_I(x)
\label{eq:L-total}
\end{eqnarray}
where ${\it L}_0(x)$ is the usual free Lagrangians with physical masses
$m_N$ for the nucleon field $\psi_N$ and $m_\pi$ for the pion field 
$\phi_\pi$, and
\begin{eqnarray}
{\it L}_I(x) =
 \bar{\psi}_N(x)\Gamma_{N,\pi N}\psi_N(x) \phi_\pi(x),.
\label{eq:L-int}
\end{eqnarray}
Here $\Gamma_{N,\pi N}$ denotes the  physical $\pi NN$ coupling
($ \sim f_{\pi NN}$). It is not the bare coupling $\Gamma_0$ in 
Eq.~(\ref{eq:barel}). The Hamiltonian density ${\it H}(x)$ can be derived from
Eqs.~(\ref{eq:L-total})-(\ref{eq:L-int}) by using the 
standard method of canonical
quantization. We then define the Hamiltonian as
\begin{eqnarray}
H = \int  {\it H}(\vec{x}, t=0)\,d\vec{x} \,.
\end{eqnarray}
The resulting Hamiltonian can be written as
\begin{eqnarray}
H=H_0 + H_I
\label{eq:htot}
\end{eqnarray}
with 
\begin{eqnarray}
H_0 &=&\int d\vec{k} \,[ E_N(k) b^\dagger_{\vec{k}}b_{\vec{k}} + 
E_\pi(k)  a^\dagger_{\vec{k}}a_{\vec{k}}]
\label{eq:h0} \\
H_I&=& \Gamma_{N\leftrightarrow \pi N} \nonumber \\
   &=&\int d\vec{k}_1 d\vec{k}_2 d\vec{k} \,
\delta (\vec{k}-\vec{k}_1-\vec{k}_1 )[ (\Gamma_{N,\pi N}(\vec{k}_1-\vec{k}_2)
 b^\dagger_{\vec{k}}  b_{\vec{k}_1} a_{\vec{k}_2})+(c.c.)]
\label{eq:hint}
\end{eqnarray}
where $b^\dagger$ and $a^\dagger$ ($b$ and $a$)
are the creation (annihilation) operators for the nucleon and
the pion, respectively. For simplicity, we drop the terms involving
the  anti-nucleon operator $d^+$ and $d$.
Note that $H$ along with the other constructed generators $\vec{P}$, 
$\vec{K}$, and $\vec{J}$ define the instant-form 
relativistic quantum mechanical
description of $\pi N$ scattering. We will work in the center of mass frame
and hence the forms of these other
generators of Lorentz group are not
relevant in the following derivations.

The essence of the unitary transformation method is to extract 
an effective Hamiltonian in a `few-body' space defined by an
unitary operator $U$, such that the resulting scattering equations can be
solved in practice.
Instead of the original equation of motion 
$H|\alpha\rangle  = E_\alpha |{\alpha}\rangle $, we consider
\begin{eqnarray}
H^\prime |\bar{\alpha}\rangle  = E_\alpha |\bar{\alpha}\rangle 
\end{eqnarray}
where
\begin{eqnarray}
H^\prime &=& UHU^+\nonumber \\
|\bar{\alpha}\rangle  &=& U |\alpha\rangle 
\end{eqnarray}
In the  approach of Sato, Kobayashi and 
Ohtsubo~\cite{sko} (SKO), 
the first step is to decompose the
interaction Hamiltonian $H_I$, Eq.~(\ref{eq:hint}), into two parts
\begin{eqnarray}
H_I &= &H_I^P + H_I^Q
\label{eq:hpq}
\end{eqnarray}
where $H_I^P$ defines the process $a \rightarrow bc$ with 
$m_a\geq m_b+m_c$ which can take place in the free space, and
$H_I^Q$ defines the virtual process with $m_a < m_b+m_c$.
For the simple interaction Hamiltonian, Eq.~(\ref{eq:hint}), it 
is clear that $H^P_I=0$ and $H^Q_I= H_I$.

The essence of the SKO method is to  eliminate the virtual processes 
from transformed Hamiltonian $H^\prime$ by choosing an appropriate 
unitary transformations $U$. This can be done systematically by using 
a perturbative expansion of $U$ in powers of coupling constants. As 
a result the effects of 'virtual processes' are included in the
effective operators in the transformed Hamiltonian.

Defining $U = \exp(-iS)$ by a hermitian operator S and expanding
$U = 1-iS + ...$\,, the transformed Hamiltonian can be written as
\begin{eqnarray}
H' &=& UHU^+  \nonumber \\
&=& U(H_0 + H^P_I + H^Q_I ) U^+  \nonumber \\
&=& H_0 + H^{P}_I + H^{Q}_I + [H_0 , iS\,] 
+ [H_I , iS\,] + {1\over 2!} \, \Big[ [H_0 , iS\,] , iS\,\Big] + \cdots\,.
\label{eq:hpq1}
\end{eqnarray}
To eliminate from Eq.~(\ref{eq:hpq1}) the virtual processes which are 
of first-order in the coupling constant,  the SKO method imposes the 
condition that
\begin{equation}
H^{Q}_{I} + [H_0 , iS\,] = 0 \,.
\label{eq:hpq-s}
\end{equation}
Since $H_0$ is a diagonal operator in Fock-space, Eq.~(\ref{eq:hpq-s}) 
implies that $iS$ must have the same operator structure of $H^{Q}_{I}$ 
and is of first order in the coupling constant. By using Eq.~(\ref{eq:hpq-s}), 
Eq.~(\ref{eq:hpq1}) can be written as
\begin{eqnarray}
H^\prime = H_0 + H_I^\prime \,, 
\end{eqnarray}
with
\begin{eqnarray}
H_I^\prime = H_I^P + [H_I^P,iS\,] + \frac{1}{2} \,[H_I^Q,iS\,]
+ \mbox{higher order terms}\,.
\label{eq:hpq2}
\end{eqnarray}
Since $H_I^P$, $H_I^Q$, and $S$ are 
all of the first order in the coupling constant, 
all processes included in the second and third terms of
the $H_I^\prime$ are of the second order in coupling constants.

We now turn to illustrating how the constructed $H_I^\prime$ of
Eq.~(\ref{eq:hpq2}) can be used to describe the $\pi N$ scattering
if the higher order terms are dropped. We consider
the simple Hamiltonian defined by Eqs.~(\ref{eq:htot})-(\ref{eq:hint}) which
gives $H^P_I=0$ and $H^Q_I=\Gamma_{N\leftrightarrow \pi N}$.
Our first task is to find $S$ by solving Eq.~(\ref{eq:hpq-s}) within the
Fock space spanned by the eigenstates of $H_0$
\begin{eqnarray}
& &H_0|N\rangle  = m_N|N\rangle  \\
& &H_0|\vec{k},\vec{p} \rangle  = (E_\pi(k)+E_N(p))|\vec{k},\vec{p}\rangle  \\
& &H_0 | \vec{k}_1,\vec{k}_2,\vec{p} \rangle  = 
((E_\pi(k_1)+E_\pi(k_2)+ E_N(p)) | \vec{k}_1,\vec{k}_2,\vec{p} \rangle  \\
& & \cdots \,.\nonumber
\end{eqnarray}
For two eigenstates $f$ and $i$ of $H_0$, the solution
of Eq.~(\ref{eq:hpq-s}) clearly is
\begin{eqnarray}
\langle f|(iS)|i\rangle  = \frac{-\langle f|H^Q_I|i\rangle }{E_f- E_i} \,.
\end{eqnarray}
For the considered $H^Q_I=\Gamma_{N\leftrightarrow \pi N}$
we thus get the following non-vanishing matrix elements
\begin{eqnarray}
\langle \vec{k}\vec{p}|(iS)|N\rangle  &=& 
\Gamma_{N,\pi N}(k) 
\,\frac{-1}{E_\pi(k)+E_N(p)-m_N}
\,\delta(\vec{k}+\vec{p})
\label{eq:r-1} \\
\langle N|(iS)|\vec{k}^{\,\prime}\vec{p}^{\,\prime}\rangle  &=& 
\frac{-1}{m_N-E_\pi(k')-E_N(p')}
\,\Gamma^*_{N,\pi N}(\vec{k}^{\,\prime})
\,\delta(\vec{k}^{\,\prime}+\vec{p}^{\,\prime})\label{eq:r-2}  \\
\end{eqnarray}
and
\begin{eqnarray}
\langle \vec{k}_1,\vec{k}_2,\vec{p}_I|(iS)|\vec{k}^\prime \vec{p}^{\,\prime}\rangle  
&=&\Gamma^*_{N,\pi N}(k_1)\frac{-\delta(\vec{k}'-\vec{k}_2)
\delta(\vec{p}^{\,\prime}-\vec{k}_1-\vec{p}_I)}
{E_\pi(k_1)+E_\pi(k_2)+E_N(p_I)-E_\pi(k')
-E_N(p')} \label{eq:r-3}\nonumber \\
&=&\Gamma^*_{N,\pi N}(k_1)\frac{-\delta(\vec{k}'-\vec{k}_2)\delta(\vec{p}^{\,\prime}-\vec{k}_1-\vec{p}_I)}
{E_\pi(k_1)+E_N(p_I)-E_N(p')}
\label{eq:r-4} \\
& & \nonumber \\
\langle \vec{k}\vec{p}|(iS)|\vec{k}_1,\vec{k}_2,\vec{p}_I\rangle  
&=&\Gamma_{N,\pi N}(k_2)\frac{-\delta(\vec{k}-\vec{k}_1)\delta(\vec{p}^{\,\prime}-\vec{k}_2-\vec{p}_I)}
{E_\pi(k)+E_N(p)-E_\pi(k_1)-E_\pi(k_2)
-E_N(P_I)} \nonumber \\
&=&\Gamma_{N,\pi N}(k_2)\frac{-\delta(\vec{k}-\vec{k}_1)\delta(\vec{p}^{\,\prime}-\vec{k}_2-\vec{p}_I)}
{E_N(p)-E_\pi(k_2)-E_N(p_I)} \label{eq:r-5}
\end{eqnarray}
With the above matrix elements and recalling that $H^P_I=0$ and 
$H^Q_I=\Gamma_{N\leftrightarrow \pi N}$ for
the considered simple case, the matrix element of the effective Hamiltonian
Eq.~(\ref{eq:hpq2}) in the center of mass frame ($\vec{p}=-\vec{k}$
and $\vec{p}^{\,\prime}=-\vec{k}'$) is
\begin{eqnarray}
\langle \vec{k}| H_I^\prime |\vec{k}'\rangle  &=&
\frac{1}{2}\sum_{I}[(\langle \vec{k}|\Gamma_{N\leftrightarrow \pi N}|I\rangle \langle I|
(iS)|\vec{k}'\rangle - \langle \vec{k}|(iS)|I\rangle \langle I|\Gamma_{N\leftrightarrow \pi N}|\vec{k}'\rangle ]
\end{eqnarray}
The only possible intermediate states are 
$|I\rangle  = |N\rangle  + |\pi(k_1)\pi(k_2) N(P_I)\rangle $. By using
 Eqs.~(\ref{eq:r-1})-(\ref{eq:r-5}) we then obtain
\begin{eqnarray}
\langle \vec{k}| H_I^\prime |\vec{k}'\rangle  &=&
 v^{(s)}(\vec{k},\vec{k}')+v^{(u)}(\vec{k},\vec{k}')
\label{eq:veff}
\end{eqnarray}
where
\begin{eqnarray}
v^{(s)}(\vec{k}, \vec{k}')&=&\frac{1}{2}\Gamma^*_{N,\pi N}(k)
\left [\frac{1}{E_\pi(k)+E_N(k)-m_N} + \frac{1}{E_\pi(k')+E_N(k')-m_N}\right]
\Gamma^*_{N,\pi N}(k')
\label{eq:veff-s} \\
v^{(u)}(\vec{k}, \vec{k}')&=&\frac{1}{2}\Gamma^*_{N,\pi N}(k')
\left[\frac{1}{E_N(k)-E_\pi(k')-E_N(\vec{k}+\vec{k}')} \right. \nonumber \\
&+&\left. \frac{1}{E_N(k')-E_\pi(k)-E_N(\vec{k}+\vec{k}')}\right]\Gamma_{N,\pi N}(k)
\label{eq:veff-u} 
\end{eqnarray}
Note that $v^{(s)}$ of Eq.~(\ref{eq:veff-s}) is due to the intermediate
'physical' nucleon state state $|I\rangle  = |N\rangle $. Here we see an
important difference between  $v^{(s)}$  and $v^{(pole)}$ of
Eq.~(\ref{eq:v-pole}) for the nucleon-pole term which is due to a 
bare nucleon state within the 3dBS models. There is no bare mass $m^0_N$
and energy-dependence in $v^{(s)}$. This is a consequence of the unitary
transformation which eliminates the 'virtual' $\pi N \leftrightarrow N$ process. 

With the above derivations, the effective Hamiltonian Eq.~(\ref{eq:hpq2}) 
can be explicitly written as
\begin{eqnarray}
H'= H_0 + V
\label{eq:effh-0}
\end{eqnarray}
where
\begin{eqnarray}
H_0 &=&\int d\vec{k} \,[ E_N(k) b^\dagger_{\vec{k}}b_{\vec{k}} +
E_\pi(k)  a^\dagger_{\vec{k}}a_{\vec{k}}]
\label{eq:eff-h0} \\
V &=& \int d\vec{k}d\vec{k}' \, [v^{(s)}(\vec{k}, \vec{k}')+v^{(u)}(\vec{k}, \vec{k}')]
a^\dagger_{\vec{k}}b^\dagger_{-\vec{k}}
a_{\vec{k}'}
b_{-\vec{k}'} \label{eq:veff-0}
\end{eqnarray}

To see further the difference
between the models from the unitary transformation method and the
3dBS method, let us first recall
how the bound states and resonances are defined in a Hamiltonian formulation.
In operator form the reaction amplitude for a Hamiltonian Eq.~(\ref{eq:effh-0})
is defined by
\begin{eqnarray}
t(E) = V + V\frac{1}{E-H_0+i\epsilon}t(E) \label{eq:lseq}
\end{eqnarray}
or
\begin{eqnarray}
t(E) = V + V\frac{1}{E-H'+i\epsilon}V \label{eq:loweq}\,.
\end{eqnarray}
The analytic structure of scattering amplitude can be most transparently
seen by using the spectral expansion of the Low equation~(\ref{eq:loweq})
\begin{eqnarray}
\langle k'|t(E)|k\rangle  &=&
 \langle k'|V|k\rangle + \sum_{i} \frac{\langle k'|V|\Phi_{\epsilon_i}\rangle \langle \Phi_{\epsilon_i}|V|k\rangle }
{E - \epsilon_i} \nonumber \\
&+& 
\int_{E_{th}}^{\infty} dE'\,\frac{\langle k'|V|\Psi^{(+)}_{E'}\rangle \langle \Psi^{(+)}_{E'}|V|k\rangle }
{E-E'+i\epsilon} \nonumber \\
\label{eq:c1}
\end{eqnarray}
where  $E_{th}$ is the threshold of the
reaction channels,  $\Phi_{\epsilon_i}$ and $\Psi^{(+)}_{E'}$ are the discrete
bound states and the scattering states, respectively.
They form a complete set and satisfy
\begin{eqnarray}
H'|\Phi_{\epsilon_i}\rangle  &=& \epsilon |\Phi_{\epsilon_i}\rangle  \label{eq:heff-bs} \\
H'\Psi^{(+)}_{E'}\rangle  &=& E' |\Psi^{(+)}_{E'}\rangle  \label{eq:heff-scatt}
\end{eqnarray}
Of course bound state energies $\epsilon_i$ are below the production threshold
$E_{th}$. We now note that due to the two-body nature of $V$ defined by 
Eq.~(\ref{eq:veff-0}), Eq.~(\ref{eq:heff-bs}) has the one-nucleon solution
$H'|N\rangle  = H_0|N\rangle  = m_N |N\rangle $. But it does not contribute to the
second term of Eq.~(\ref{eq:c1}) because $\langle \pi N|V|N\rangle  =0$.
Thus the amplitude Eq.~(\ref{eq:c1}) does not have a nucleon pole which 
corresponds to a bound state with the mass of the physical nucleon and is formed
by the $physical$ $N$ and $\pi$ of the staring Lagrangian
Eq.~(\ref{eq:L-total}). This is consistent with the experiment. 

Here we note that in the Hamiltonian formulation, the amplitude 
$\langle k'|t(E)|k\rangle $  depends on three independent variables: 
energy $E$ and momenta $k$ and $k'$. We can analytically continue 
this amplitude to complex $E$-plane for any $k$ and $k'$. In the 
complex $E$-plane, the bound state poles and unitarity cuts are 
on the  real axis of the physical sheet and the resonance poles 
are on the unphysical sheet. These analytic properties with 
respect to the energy variable $E$ are independent of the 
momentum variables $k$ and $k'$. As mentioned at the end 
of section II, the on-shell matrix element $\langle k_0|t(E)|k_0\rangle $ could
have poles from the form factors which are needed to regularize the 
potential $V$. Thus the analytic structure of $\langle k_0|t(E)|k_0\rangle $
can be different from that of the approaches based on dispersion relations.

To further see the differences with the 3dBS models and the approaches 
based on  dispersion relations, let us solve Eq.~(\ref{eq:lseq}) by 
considering only $V= v^{(s)}$. The matrix element of the
scattering equation defined by Eq.~(\ref{eq:lseq}) is identical to
Eqs.~(\ref{eq:3dBS-eq})-(\ref{eq:3dBS-eq-p}).
With the separable form  Eq.~(\ref{eq:veff-s}) of $V=v^{(s)}$, the above 
equations can be solved explicitly. The solution is
\begin{eqnarray}
T(k,k',E) & = & \frac{N(k,k',E)}{D(E)}
\label{eq:t-sol}
\end{eqnarray}
with
\begin{eqnarray}
D(E) & = & (1 - D_1(E))^2 - D_0(E)D_2(E)
\label{eq:bd} \\
& & \nonumber \\
N(k,k',E)&=&
 \Gamma^*_{N,\pi N}(k')\left[
\frac{D_0(E)}{4}\frac{1}{(E(k)-m)(E(k')-m_N)} \right.\nonumber \\
&+& \left. \frac{(1-D_1(E))}{2} \left(\frac{1}{E(k)- m} +\frac{1}{E(k')-m_N}\right)
+D_2(E)\right]\Gamma_{N,\pi N}(k)
\label{eq:t-n}
\end{eqnarray}
where $E(k)=E_\pi(k)+E_N(k)$ and
\begin{eqnarray}
D_0(E) & = & \int_0^\infty q^2dq \;|\Gamma_{N,\pi N}(q)|^2 \frac{1}{E - E(q)+ i\epsilon}
\label{eq:bd0}\\
D_1(E) & = & \int_0^\infty q^2dq\; |\Gamma_{N,\pi N}(q)|^2
        \frac{1}{2(E - E(q)+ i\epsilon)(E(q) - m)}\label{eq:bd1}\\
D_2(E) & = & \int_0^\infty q^2 dq\; |\Gamma_{N,\pi N}(q)|^2
        \frac{1}{4(E - E(q)+ i\epsilon)(E(q) - m)^2}\label{eq:bd2}
\end{eqnarray}
With some inspection, one can see that the amplitude Eq.~(\ref{eq:t-sol})
does not have the nucleon pole condition $D(E=m_N)  =0$ for any $k$ and $k'$. 
This is what one expects from the spectral expansion Eq.~(\ref{eq:c1}). 
If we take the on-shell matrix element $E=E(k_0)=E(k)=E(k')$, one then finds
\begin{eqnarray}
t(k_0,k_0,E) & = &
 \Gamma^*_{\pi NN}(k_0)
\left[\frac{d_1(E)}{E-m}+ \frac{d_0(E)}{4(E-m^2) ^2} + d_2(E)\right]\Gamma_{\pi NN}(k_0)
\label{eq:ont}
\end{eqnarray}
where
where
\begin{eqnarray}
d_0(E) & = & \frac{D_0(E)}{D(E)} \,,\label{eq:d0}\\
d_1(E) & = & \frac{1 - D_1(E)}{D(E)} \,, \label{eq:d1}\\
d_2(E) & = & \frac{D_2(E)}{D(E)} \,.\label{eq:d2}
\end{eqnarray}
The first term in the right-hand side of Eq.~(\ref{eq:ont}) does have
a pole at $E=M_N$ of the dispersion relations. But it has additional
double poles from the second term as well as from the $\pi NN$ form 
factor  which is often parameterized as a dipole form 
$ \Gamma_{\pi NN}(k)=(\Lambda^2/(\Lambda^2+k^2))^2$. It should be 
noted that a pole corresponding to a bound state in a Hamiltonian 
formulation must be for arbitrary $k$ and $k'$. Thus the pole 
only from the  on-shell matrix element Eq.~(\ref{eq:ont}) is 
not a $\pi N$ bound state with mass $m_N$. Eq.~(\ref{eq:ont}) shows again 
that a dynamical model deduced from relativistic quantum field theory
does not have, and is not required to have, the same analytic structure
of the amplitudes in the approaches based on
dispersion relations in the S-matrix theory. 

To end this section, let us mention that the unitarity condition
only requires that an acceptable model must have a unitarity cut 
in the physical region $E \geq m_\pi + m_N$. This is trivially satisfied 
in the model defined by the effective Hamiltonian 
Eqs.~(\ref{eq:eff-h0})-(\ref{eq:veff-0}) since the interaction $V$ 
is energy independent. This is an important advantage of applying 
the method of unitary transformation to develop a multi-channels 
multi-resonances reaction models for investigating meson-nucleon 
reactions in the nucleon resonance region, as developed in 
Ref.~\cite{msl07}. In a model with an energy-dependent $V$ such 
as the AAY model the unitarity condition is much more difficult to 
satisfy, and the analytic continuation of the scattering $t$-matrix 
defined by Eqs.~(\ref{eq:c1}) to complex $E$-plane is in general 
much more involved.

\section{Time-ordered perturbation theory}
Treating the Hamiltonian Eq.~(\ref{eq:htot}) in time-ordered perturbation
theory~\cite{schweber}, the matrix elements of the transition operator
can be represented by a series expansion defined by all diagrams 
containing an incoming and outgoing pion-nucleon state
\begin{eqnarray}
\langle \pi N|t(E)|\pi N\rangle  &=& \langle \pi N|H_I\frac{1}{E-H_0+i\epsilon}H_I|\pi N\rangle 
\nonumber \\
&+&
\langle \pi N|H_I\frac{1}{E-H_0+i\epsilon}H_I\frac{1}{E-H_0+i\epsilon}H_I
\frac{1}{E-H_0+i\epsilon}H_I|\pi N\rangle  \nonumber \\
&+& \cdots\,.
\end{eqnarray}
In the approach of the Julich group~\cite{julich90,julich00}, 
the partial sum of this series is written as a three-dimensional
integral equation which takes the same form of Eqs.~(\ref{eq:3dBS-eq})-(\ref{eq:3dBS-eq-p}).
In the simple model defined by Eq.~(\ref{eq:h0})-(\ref{eq:hint}), the resulting
potential can be schematically written as (dropping anti-nucleon terms)
\begin{eqnarray}
v(\vec{k},\vec{k}',E)=v^{(s)}\vec{k},\vec{k}',E)+v^{(u)}\vec{k},\vec{k}',E)
\label{eq:veffj}
\end{eqnarray}
where
\begin{eqnarray}
& &v^{(s)}(\vec{k},\vec{k}',E) =\Gamma^*_{N,\pi N}(k)
\;\frac{1}{E-m^0_N}\;\Gamma_{N,\pi N}(k') \label{eq:veffj-s} \\
& &v^{(u)}(\vec{k},\vec{k}',E) =\Gamma^*_{N,\pi N}(k')
\left[\frac{1}{E-E_N(\vec{k}+\vec{k'})-E_\pi(k)-E_\pi(k')}\right]
\Gamma_{N,\pi N}(k) \label{eq:veffj-u}
\end{eqnarray}
As can be readily seen the term $v^{u}(E)$ has the singularity of the  
$\pi \pi N$ cut, and they depart from the starting Hamiltonian 
Eq.~(\ref{eq:h0})-(\ref{eq:hint}) by using the bare mass $m^0_N$ to define the
s-channel term $v^{(s)}$.
The above potential has the same form of the matrix element of that defined by 
Eqs.~(\ref{eq:3dbs-v})-(\ref{eq:v-pole}).
Thus the interpretation of their analysis of nucleon pole term is similar
to what described in section II.

\section{Summary}
In this  paper, we have examined three methods for constructing
meson-nucleon reaction models from relativistic quantum field theory.
For the models based on the
three-dimensional reductions of Bethe-Salpeter equation
 and the time-ordered perturbation theory, 
the driving terms of the resulting three-dimensional scattering equations
in general contain a nucleon pole term determined by a bare nucleon $N_0$.
We show that the commonly used procedure of imposing the nucleon pole 
condition to fix the bare nucleon parameters is related to the
assumption that the nucleon is a bound state made of
a bare core $N_0$ and meson cloud. To correctly implement this
nucleon substructure into the scattering equation, it is necessary to
consider $\pi \pi N$ unitarity condition as achieved within the model
of Aaron, Amado and Young~\cite{aay}. 

We have  given a pedagogical and explicit explanation of 
the method of unitary transformation which has been applied in recent
years to investigate 
meson-nucleon reactions~\cite{sl96,msl07,jlms07,jlmss08,kjlms08}. 
Since only $physical$ nucleons and pions are the basic degrees of freedom of
the derived effective Hamiltonian, the resulting $\pi N$ amplitude
does not have a nucleon pole at $E=M_N$ in the complex $E$-plane.
This is due to the fact that the one-nucleon problem is decoupled from the
two-particle problem by the unitary transformation and
the resulting $\pi N$ potential is energy independent. There is no $\pi N$
interaction due to the propagation of a $bare$ nucleon in this formulation.
We explain  how this can be understood 
from the general principles of a Hamiltonian formulation of reactions.

We also show that the scattering amplitudes from these three methods
do not have the same analytic structure of the amplitude from the
approach based on the dispersion relations of the S-matrix theory. 
Even one imposes the nucleon pole
condition to make the connection to the S-matrix theory, 
the constructed 
models can have poles from the form factors which are needed
to regularize the  potentials for solving the resulting
scattering equations. We emphasize that there is no compelling and
rigorous theoretical argument to request that the constructed models should
have the analytic structure of the dispersion relations.
There exits no rigorous derivation of the dispersion relations of S-matrix 
theory 
from relativistic quantum field theory~\cite{gw}. 
Historically, the S-matrix theory is considered as an alternative to
 relativistic quantum field theory to study strong interactions.
Either one of them is a good starting point for developing
phenomenological models for analyzing the data. For investigating
 multi-channels and multi-resonances reactions, the models deduced
from relativistic field theory appear to be more practical.

\begin{acknowledgments}
We thank C. Hanhart for bringing our attention to the double pole 
of the on-shell amplitude Eq.~(\ref{eq:ont}) in the model based on
the method of unitary transformation.
This work is supported by the U.S. Department of Energy, 
Office of Nuclear Physics Division, under contract No. DE-AC02-06CH11357, 
and Contract No. DE-AC05-060R23177 under which Jefferson Science Associates 
operates Jefferson Lab, and by the Japan Society for the Promotion of Science,
Grant-in-Aid for Scientific Research(c) 20540270. This work is also
partially supported by Grant No. FIS2008-01661/FIS from MCIIN and 
CPAN CSD2007-00042 Consolider Ingenio 2010 (Spain). 
\end{acknowledgments}

{99}

\end{document}